\documentclass[prd,showpacs,amsmath,amssymb,superscriptaddress,preprintnumbers,nofootinbib,showkeys]{revtex4}
\usepackage{setspace}
\usepackage{graphicx}
\usepackage{float}
\usepackage{caption}
\begin{document}

\newcommand{\nablab}{{\mathop {\rule{0pt}{0pt}{\nabla}}\limits^{\bot}}\rule{0pt}{0pt}}

\title{Kinetics of relativistic axionically active plasma in the field of dynamic aether: \\ Part II: Particle dynamics and symptoms of plasma instability}

\author{Alexander B. Balakin and Kamil R. Valiullin}
\email{Alexander.Balakin@kpfu.ru}
\email{kr_valiullin@mail.ru}
\affiliation{Department of
General Relativity and Gravitation, Institute of Physics, Kazan
Federal University, Kremlevskaya str. 18, Kazan 420008,
Russia}

\date{\today}

\begin{abstract}
In the paper , \cite{I} the first part of our work, we established the model of interaction of five elements: first, a unit time-like vector field, associated with the velocity of dynamic aether, second, the pseudoscalar field describing an axionic component of the dark matter, third, the electromagnetic field, fourth, the relativistic multi-component plasma and, fifth, the gravitational field. The initial step of analysis of this model was
the reconstruction of the equilibrium states of the axionically-aetherically active plasma, which are characterized by frequent inter-particle collisions. In the second part of the work, presented here, in the framework of isotropic homogeneous cosmological model we consider the collisionless limit of the theory, and thus, we focus on the analysis of the particle dynamics under the influence of axionically-aetherically induced forces. The complete distribution functions, as exact solutions to the kinetic equations, are presented. Based on five examples of the background solutions of the model, we discuss the symptoms of plasma instability provoked by the influence of forces of the friction and tidal types.
\end{abstract}
\maketitle

\section*{Introduction}

In the paper \cite{I} we have formulated the complete formalism of the extended version of the kinetic theory of the relativistic axionically active multi-component plasma, which is based on inclusion of a unit time-like vector field, associated with the velocity of dynamic aether \cite{A2,A3}, into the scheme of interactions. In the first part of the work we focused on the analysis of equilibrium states of the plasma, and thus the distribution functions of this multi-component plasma were considered to be of the J\"uttner-Chernikov form \cite{RKT}. When the collisionless plasma is under consideration, the  corresponding distribution functions can be found as arbitrary functions of seven first integrals of the characteristic equations (integrals of motion) attributed to the kinetic equations. Search for these integrals of motion is associated with the analysis of the plasma particle motion, that is why in this work we focus on the details of the particle dynamics in the framework of the \textit{Friedmann-Lema$\hat{\iota}$tre-Robertson-Walker} (FLRW) cosmological model.

The paper is organized as follows. Section II contains the description of the mathematical formalism reduced to the FLRW cosmological model. In section III we analyze the solutions to the equations of dynamics of plasma particles. Section IV contains conclusions.

\section{The Formalism}
\subsection{Background field configuration}

The model, which we consider in this work, has a hierarchical structure. The gravity field, dynamic aether, dark matter and dark energy form a basic global field/matter configuration. The kinetic system is considered to be a test subsystem, i.e., it evolves on the given space-time background.
The background model has the FLRW type metric
\begin{equation}
ds^2 = c^2 dt^2 - a^2(t) \left[(dx^1)^2 + (dx^2)^2 +(dx^3)^2 \right] \,.
\label{M1}
\end{equation}
Respectively, the aether velocity four-vector is of the form $U^j = \delta^j_0$, and the covariant derivative $\nabla_k U_j$ is symmetric
\begin{equation}
\nabla_k U_j = H (g_{kj}-U_k U_j) \,.
\label{M2}
\end{equation}
Here and below $H {=}\frac{\dot{a}}{a}$ is the Hubble function; the dot denotes differentiation with respect to time. For this model the acceleration four-vector vanishes, $D U_j {=} U^k \nabla_k U_j {=} 0$, the antisymmetric vorticity tensor and symmetric traceless shear tensor are equal to zero, $\omega_{kj}{=}0$, $\sigma_{kj}{=}0$. Only the scalar of expansion of the aether flow $\Theta {=} \nabla_k U^k {=} 3 H $ is non-vanishing.

The axion field $\phi$ is the function of time only; it satisfies the equation (see \cite{I})
\begin{equation}
(1+ {\cal A})\left[\ddot{\phi} + 3H \dot{\phi} \right] + \dot{{\cal A}} \dot{\phi} + m^2_A \sin{\phi} = 0 \,.
\label{M55}
\end{equation}
Since $\phi$ depends on the cosmological time only, we can use the simplification $\nabla_k \phi = U_k \dot{\phi}$.

Concerning the solutions to the gravity field equations, we can mention the following details.
There are few versions of the scale factor $a(t)$, but the most known solutions can be presented in the exponential (de Sitter-like) form
\begin{equation}
a(t) = a(t_0) e^{H_{0}(t-t_0)} \,,
\label{M3}
\end{equation}
and in the power-law form
\begin{equation}
a(t) = a(t_0) \left(\frac{t}{t_0}\right)^{\gamma} \,.
\label{M21}
\end{equation}
As  it was shown in the series of works \cite{E1,E2,E3,E4,E5},
the following three exact solutions also are interesting for applications: first,
the anti-Gaussian solution describing the so-called symmetric bounce
\begin{equation}
a(t) = a(t_*) e^{h t^2} \,,
\label{M5}
\end{equation}
second, the super-exponential solution describing the so-called super inflation
\begin{equation}
a(t) = a(t_0) e^{\nu \sinh{H_{*} (t-t_0)}} \,,
\label{M4}
\end{equation}
and the quasi-periodic solution associated with the so-called reheating in the early Universe
\begin{equation}
a(t) = a(t_0) e^{\rho \sin{\Omega(t-t_0)}} \,.
\label{M6}
\end{equation}
The model parameters $h$, $\nu$, $\rho$, $\Omega$, $\gamma$, $H_{*}$ and their physical sense can be found in the quoted papers \cite{E1,E2,E3,E4,E5}. For instance, the solutions of the super-exponential and quasi-periodic types appear in the model, when the dark energy and non-axionic dark matter are the rheological cosmic substrates and their interaction is described by the integral operators of the Volterra type \cite{E3,E4}.

We also assume that the background electromagnetic field is absent providing the space - time to be isotropic and homogeneous; the cooperative electromagnetic field in plasma is the test one, it does not form an essential contribution to the total stress-energy tensor of the system.

\subsection{Kinetic equations for the collisionless plasma, equations of characteristics and  generalized forces}

\subsubsection{Characteristic equations}

The formalism of kinetic theory is based on a set of relativistic kinetic equations; in the collisionless plasma they have the form
\begin{equation}
\frac{p^j}{m_{(a)} c} \left(\frac{\partial }{\partial x^j} - \Gamma^l_{jk} p^k \frac{\partial }{\partial p^l}\right)f_{(a)} + \frac{\partial}{\partial p^i} \left[{\cal F}_{(a)}^i f_{(a)} \right] = 0\,.
\label{9}
\end{equation}
The distribution functions $f_{(a)}(x^j, p^k)$ are marked by the index of particle sort  $(a)$; they are functions of coordinates $x^j$ and of the particle momenta $p^k$, which have the status of random variables. The terms ${\cal F}_{(a)}^i$ relate to the force four-vectors, which act on the particles of the sort $(a)$.  Characteristics for the differential equations (\ref{9}) can be written as follows:
\begin{equation}
\frac{{\cal D} p^j}{ds} = {\cal F}^j_{(a)} \,, \quad \frac{{\cal D} p^j}{ds} \equiv \frac{d p^j}{ds} + \Gamma^j_{lk} p^l \frac{p^k}{m_{(a)}c} \,.
\label{10}
\end{equation}
We assume that the particle momenta are normalized, i.e., $g_{kj} p^k p^j = m^2_{(a)}c^2$, and the rest masses of particles, $m_{(a)}$, conserve, so that
\begin{equation}
p_j \frac{{\cal D} p^j}{ds} = \frac12 \frac{{\cal D} (p_jp^j)}{ds} = \frac12 \frac{d [m^2_{(a)}c^2] }{ds} =0 \ \Rightarrow  p_j {\cal F}^j_{(a)} =0 \,.
\label{11}
\end{equation}
In other words, we consider only the forces,  which are orthogonal to the particle momentum four-vector. Also, we keep in mind that the square of particle momenta is the quadratic first integral of the characteristic equations (\ref{10}), $g_{kj} p^k p^j = K_0 = m^2_{(a)}c^2$.

\subsubsection{Generalized forces}

In the first part of work \cite{I} we have classified the forces, which can appear in the context of axionically-aetherically active plasma. Only few of these mentioned forces can appear in the isotropic homogeneous FLRW type model with the metric (\ref{M1}). Let us consider them in more details.

I. The forces of the Stokes type.

These forces are quadratic in the particle momentum and linear in the aether velocity four - vector. The relativistic friction force can be presented as follows:
\begin{equation}
{\cal F}^i_{(S)} = \frac{\lambda_{(S)}}{m_{(a)} c^2}[\delta^i_k (p^l p_l) - p^ip_k] \ U^k \ \left[1 + \nu_{(S)}\cos{\phi} \right]  \,.
\label{13}
\end{equation}
The force vanishes, when the particle four-velocity $\frac{p^j}{m_{(a)}c}$ coincides with the aether velocity $U^j$.  When $|\nu_{(S)}|>1$, the multiplier $\left[1 {+} \nu_{(S)}\cos{\phi} \right]$ behaves as  axionic switch, providing the force to be of the friction or antifriction type depending on the state of the axion field.

The second Stokes type force appears initially as the gradient-type force proportional to the gradient of the axion field $\nabla_k \phi = U_k \dot{\phi}$; it looks like the first Stokes type force.
\begin{equation}
{\cal F}^i_{(G)} = \frac{\lambda_{(G)}}{m_{(a)} c^2} \ [g^{ik} (p^l p_l) - p^i p^k]  \ U_k \dot{\phi} \ \sin{\phi}  \left[1 + \nu_{(G)}\cos{\phi} \right] \,.
\label{16}
\end{equation}

II. Tidal  forces

We indicate the forces linear in curvature as the tidal forces
\begin{equation}
{\cal F}^i_{(T)} = \frac{\lambda_{(T)}}{m_{(a)} c^2} {\cal R}^i_{\cdot kmn} p^k U^m p^n \left[1 + \nu_{(T)}\cos{\phi} \right]\,,
\label{20}
\end{equation}
where $\cal R$ is the three-parameter tensor of non-minimal susceptibility
\begin{equation}
{\cal R}_{ikmn} = \frac12 q_1 R \left(g_{im}g_{kn}-g_{in}g_{km}\right) + q_2 \left[ R_{im} g_{kn} - R_{in} g_{km} + R_{kn} g_{im} - R_{km} g_{in} \right] + q_3 R_{ikmn} \,,
\label{21}
\end{equation}
based on the Riemann tensor $R_{ikmn}$, Ricci tensor $R_{km}$ and Ricci scalar $R$.
Also, there exist left-dual and right-dual susceptibility tensors (see the force classification presented in \cite{I})
\begin{equation}
^*{\cal R}^i_{ \cdot kmn} = \frac12 \epsilon^i_{\cdot kpq} \ {\cal R}^{pq}_{\ \ mn} \,, \quad {\cal R}^{*i}_{\  \cdot kmn} = \frac12 {\cal R}^{i}_{\cdot k pq} \epsilon^{pq}_{\ \ mn}  \,.
\label{22}
\end{equation}
However, they do not contribute to the characteristic equations due to the high symmetry of the FLRW model.

We have to stress, that the quantities $\lambda_{(S)}$,  $\lambda_{(T)}$, $\nu_{(S)}$, $\nu_{(T)}$, etc., can be considered as functions of the expansion scalar $\Theta=3H$, or as constants. In addition, we assume that they can depend on the index of the particle sort $(a)$, but for the sake of simplicity we do not write this index.

\subsubsection{Reduced characteristic equations}

We keep in mind that for $j=\alpha =1,2,3$  we obtain that
\begin{equation}
\frac{Dp_j}{ds} =\frac{d p_j}{ds}-\frac{1}{2m_{(a)}c} p^s p^l \partial_j g_{sl} = \frac{d p_j}{ds} \,,
\label{025}
\end{equation}
and consider below the spatial subsystem of equations (\ref{10}) searching for the covariant component of the spatial part of the particle momentum $p_{\alpha}$:
$$
\frac{d p_{\alpha}}{ds} = \frac{\lambda_{(S)}}{m_{(a)} c^2}[g_{\alpha k} (p^l p_l) - p_{\alpha} p_k] \ U^k \ \left[1 + \nu_{(S)}\cos{\phi} \right] +
$$
$$
+ \frac{\lambda_{(G)}}{m_{(a)} c^2} \ [g_{\alpha k} (p^l p_l) - p_{\alpha} p_k] U^k \ \dot{\phi} \ \sin{\phi}  \left[1 + \nu_{(G)}\cos{\phi} \right] +
$$
\begin{equation}
+ \frac{\lambda_{(T)}}{m_{(a)} c^2} {\cal R}_{\alpha kmn} p^k U^m p^n \left[1 + \nu_{(T)}\cos{\phi} \right]
   \,.
\label{25}
\end{equation}
From the first quadratic integral of motion $p_j p^j = m^2_{(a)} c^2$ we can obtain the component $p_0=p^0>0$ of the particle momentum:
\begin{equation}
p^0(t) = \sqrt{ m^2_{(a)} c^2 + p^2(t) } \,,
\quad p^2(t) = a^{-2}(t) \left[ (p_1)^2 + (p_2)^2 +(p_3)^2 \right] \,.
\label{31}
\end{equation}
Since the spatial directions $Ox^1$, $Ox^2$, $Ox^4$ are equivalent in the FLRW space-time, it is reasonable to consider only one equation of characteristics, say, for $\alpha=1$, and to rewrite the differentiation in the left-hand sides of this equation using the relationship
$\frac{d p_1}{ds} = \frac{p^0}{m_{(a)}c^2}\frac{d p_1}{dt}$.

Also we know that for the chosen symmetry the non-vanishing components of the Riemann tensor, Ricci tensor, and Ricci scalar are
$$
R^{01}_{\ \ 01} = R^{02}_{\ \ 02} = R^{03}_{\ \ 03} = - \frac{\ddot{a}}{a} \,, \quad R_{0101} = a \ddot{a} \,,
$$
$$
R^{12}_{\ \ 12} = R^{13}_{\ \ 13} = R^{23}_{\ \ 23} = - \left(\frac{\dot{a}}{a}\right)^2 \,, \quad R_{1212} = - a^2 {\dot{a}}^2 \,,
$$
\begin{equation}
R^0_0 = -3 \frac{\ddot{a}}{a} \,, \quad R^1_1 = R^2_2 = R^3_3 = -\left[\frac{\ddot{a}}{a} + 2 \left(\frac{\dot{a}}{a}\right)^2 \right] \,,
\quad R = -6 \left[\frac{\ddot{a}}{a} + \left(\frac{\dot{a}}{a}\right)^2 \right] \,.
\label{FLRW2}
\end{equation}
Based on these auxiliary formulas one can reconstruct the non-vanishing components of the non-minimal susceptibility tensor (\ref{21})
\begin{equation}
{\cal R}^1_{ \cdot 001}={\cal R}^2_{ \cdot 002}={\cal R}^3_{ \cdot 003}= \frac{\ddot{a}}{a} \left( 3q_1 + 4q_2 + q_3 \right) + \frac{\dot{a}^2}{a^2} \left( 3q_1 + 2q_2 \right) \,,
\label{29}
\end{equation}
\begin{equation}
{\cal R}^2_{ \cdot 121}={\cal R}^3_{ \cdot 131}={\cal R}^3_{ \cdot 232}= a\ddot{a} \left( 3q_1+2q_2 \right) + \dot{a}^2 \left( 3q_1 + 4q_2 +q_3 \right) \,.
\label{30}
\end{equation}

\section{Analysis of particle dynamics}

\subsection{First special case}

Let us consider the axion field to be frozen into one of the minima of the axion potential, $\phi = 2\pi k$ ($k$ is an integer).
In this case the key equation takes the form
\begin{equation}
 \frac{1}{p_1}\frac{d p_1}{dt} = - \lambda_{(S)}  \left[1 + \nu_{(S)} \right] +\lambda_{(T)} \left\{ \dot{H} \left( 3q_1 + 4q_2 + q_3 \right) + H^2\left( 6q_1 + 6q_2 + q_3 \right) \right\} \left[1 + \nu_{(T)}\right]            \,.
\label{R25}
\end{equation}
The multipliers $\lambda_{(S)}$ and $\lambda_{(T)}$ are considered to be functions of the expansion scalar $\Theta = 3H$. Also we assume that the parameters $q_1$, $q_2$ and $q_3$ are dimensionless. Since the right-hand side of the equation  (\ref{R25}) has to have the dimensionality of inverse time, we suggest the following structure of the  multipliers $\lambda_{(S)}$ and $\lambda_{(T)}$:
\begin{equation}
\lambda_{(S)} =
\tilde{\lambda}_{(S)} H \,, \quad \lambda_{(T)} = H^{-1}\tilde{\lambda}_{(T)} \,,
\label{L8}
\end{equation}
where the quantities $\tilde{\lambda}_{(S)}$ and $\tilde{\lambda}_{(T)}$ are dimensionless constant parameters.
The solution to the equation (\ref{R25}) is of the form
\begin{equation}
p_1(t) = p_1(t_0) \left(\frac{a(t)}{a(t_0)} \right)^{\sigma_1}  \left(\frac{H(t)}{H(t_0)} \right)^{\sigma_2} \,,
\label{L7}
\end{equation}
and seven parameters are incorporated into two guiding parameters  of the model:
\begin{equation}
\sigma_1 = - \tilde{\lambda}_{(S)}  \left[1 + \nu_{(S)}\right] +\tilde{\lambda}_{(T)} \left( 6q_1 + 6q_2 + q_3 \right) \left[1 + \nu_{(T)}\right] \,,
\label{L6}
\end{equation}
\begin{equation}
\sigma_2 = \tilde{\lambda}_{(T)} \left( 3q_1 + 4q_2 + q_3 \right) \left[1 + \nu_{(T)}\right] \,.
\label{L5}
\end{equation}
In other words, the quantity (or any quantity proportional to this one)
\begin{equation}
p_1(t) [a(t)]^{-\sigma_1} [H(t)]^{-\sigma_2}= p_1(t_0) [a(t_0)]^{-\sigma_1}  [H(t_0)]^{-\sigma_2} = K_1
\label{L4}
\end{equation}
is the integral of the characteristic equations. Respectively, the integrals of motion  $K_2$ and $K_3$ appear, when we replace the index $\alpha {=}1$ by $\alpha {=}2$ and  $\alpha {=}3$.
The set of four integrals of motion  $K_0$, $K_1$, $K_2$, $K_3$ is sufficient for construction of homogeneous distribution functions $f_{(a)}(p_j)$.
For instance, we can write the exact solutions for the distribution functions, which look like the J\"uttnet-Chernikov ones, as follows:
\begin{equation}
f_{(a)}(p_j) = A_{(a)} \exp\left\{- \frac{c}{k_B T_{(a)}} \sqrt{m^2_{(a)}c^2 + \omega^2 \left(K^2_1+K^2_2+ K^2_3 \right)}  \right\} \delta(K_0-m^2_{(a)}c^2)\,,
\label{Eq}
\end{equation}
where the temperatures of the particles of the sort $(a)$, indicated as  $T_{(a)}$, do not coincide in general case, contrarily to the case of equilibrium in the multi-component plasma. The constant $\omega^2$ can be chosen so that in the absence of gravitational field the formula (\ref{Eq}) gives the standard
J\"uttnet-Chernikov distribution function. The multipliers $A_{(a)}$ describe the normalization parameters (see, e.g., \cite{RKT} for details).

When we are interested to describe the complete distribution taking into account the dependence on spatial coordinates, we have to find three supplementary integrals of motion. In order to find them, one can write the equation
$\frac{dx^1}{ds} = \frac{g^{11}p_1}{m_{(a)} c}$, which yields
\begin{equation}
\frac{dx^1}{dt} = \frac{cp^1(t_0)}{p^0(t)} \left(\frac{a(t)}{a(t_0)} \right)^{\sigma_1-2}  \left(\frac{H(t)}{H(t_0)} \right)^{\sigma_2} \,,
\label{L3}
\end{equation}
\begin{equation}
p^0(t) = \sqrt{m^2_{(a)}c^2 + p^2(t_0) \left(\frac{a(t)}{a(t_0)} \right)^{2\sigma_1-2}  \left(\frac{H(t)}{H(t_0)} \right)^{2\sigma_2} } \,.
\label{L2}
\end{equation}
Integration of this equation gives the fifth integral of motion  in quadratures
\begin{equation}
K_4 = x^1(t) - cp^1(t_0) \int^t_{t_0} d \tau  \frac{\left(\frac{a(\tau)}{a(t_0)} \right)^{\sigma_1-2}  \left(\frac{H(\tau)}{H(t_0)} \right)^{\sigma_2}}{\sqrt{m^2_{(a)}c^2 + p^2(t_0) \left(\frac{a(\tau)}{a(t_0)} \right)^{2\sigma_1-2}  \left(\frac{H(\tau)}{H(t_0)} \right)^{2\sigma_2} }} \,.
\label{L1}
\end{equation}
The integrals of motion $K_5$, $K_6$ can be obtained similarly using the replacements $p^1(t_0) \to p^2(t_0)$, $x^1(t) \to x^2(t)$, as well as,  $p^1(t_0) \to p^3(t_0)$, $x^1(t) \to x^3(t)$. Thus, arbitrary function $f_{(a)}(K_1,K_2,K_3,K_4,K_5,K_6)\delta(K_0-m^2_{(a)}c^2) $ presents the general solution of the kinetic equation (\ref{9}) in the first special case.

\subsection{Second special case}

As an illustration, we consider now the axionically induced gradient force (\ref{16}) only.
This submodel can be described by the key characteristic equation
\begin{equation}
\frac{1}{p_1}\frac{d p_1}{dt} = - \lambda_{(G)} \ \dot{\phi} \ \sin{\phi}  \left[1 + \nu_{(G)}\cos{\phi} \right]     \,.
\label{L78}
\end{equation}
In order to keep dimensionality of the left-hand and right-hand sides of this equation we have to consider $\lambda_{(G)}$ to be constant. Integration of this equation yields
\begin{equation}
p_1(t)\exp\left\{- \lambda_{(G)}  \left[\cos{\phi}(t) + \frac12 \nu_{(G)}\cos^2{\phi}(t) \right] \right\} = p_1(t_0) \exp\left\{ - \lambda_{(G)}  \left[\cos{\phi}(t_0) + \frac12 \nu_{(G)}\cos^2{\phi}(t_0) \right] \right\}   = {\cal K}_1  \,.
\label{L87}
\end{equation}
The integrals of motion  ${\cal K}_2$ and ${\cal K}_3$ can be found similarly.
As for the integral of motion ${\cal K}_4$, we obtain
\begin{equation}
{\cal K}_4 = x^1(t) - cp^1(t_0) \int^t_{t_0} \frac{d \tau}{p^0(\tau)}  \left(\frac{a(\tau)}{a(t_0)} \right)^{-2}
\exp\left\{\lambda_{(G)}[\cos{\phi(\tau)}{-}\cos{\phi(t_0)}]\left[1{+} \frac12 \nu_{(G)}[\cos{\phi(\tau)}{+}\cos{\phi(t_0)}] \right]  \right\}\,,
\label{L109}
\end{equation}
\begin{equation}
p^0(\tau) = \sqrt{m^2_{(a)}c^2 + p^2(t_0)  \left(\frac{a(\tau)}{a(t_0)} \right)^{-2}
\exp\left\{2\lambda_{(G)}[\cos{\phi(\tau)}{-}\cos{\phi(t_0)}]\left[1{+} \frac12 \nu_{(G)}[\cos{\phi(\tau)}{+}\cos{\phi(t_0)}] \right]  \right\}
}\,.
\label{L107}
\end{equation}
The integrals of motion ${\cal K}_5$ and ${\cal K}_6$ can be found similarly. Again, general solutions to the kinetic equations have the form
$f_{(a)}({\cal K}_1,{\cal K}_2,{\cal K}_3,{\cal K}_4,{\cal K}_5,{\cal K}_6)\delta(K_0-m^2_{(a)}c^2) $.

\section{Examples of particle behavior and symptoms of plasma instability }

Now we consider two physically valuable quantities: first, the physical component of particle momentum
${\cal P}(t) = \pm \sqrt{- g^{11}p_1^2}$, and the particle energy
${\cal E}(t) = c U^j p_j = c p^0(t)$, for five space-time platforms, which can be described by the scale factors listed above.

\subsection{The case of constant axion field}

Let us start with the analysis of the physical component of the particle momentum
\begin{equation}
{\cal P}(t) = {\cal P}(t_0)  \left[\frac{a(t)}{a(t_0)}\right]^{\sigma_1-1} \left[\frac{H(t)}{H(t_0)}\right]^{\sigma_2}
\,.
\label{4021}
\end{equation}

\subsubsection{de Sitter type inflation}

For the de Sitter space-time with the scale factor and Hubble function given by
\begin{equation}
a(t)=a(t_0) e^{H_0 (t-t_0)} \,, \quad H(t)= H(t_0) = H_0 = {\rm const}
\label{41}
\end{equation}
the law of evolution of the physical momentum component has the form
\begin{equation}
{\cal P}(t) = {\cal P}(t_0)  e^{H_0(\sigma_1-1)(t-t_0)}
\,.
\label{401}
\end{equation}
The particle energy behaves as
\begin{equation}
{\cal E}(t) = c \sqrt{m^2_{(a)}c^2 + p^2(t_0) e^{2H_0 \left(\sigma_1-1\right)(t-t_0) }} \,.
\label{L29}
\end{equation}
When $\sigma_1>1$, the particle  momentum and energy  grow,  and the particle becomes ultra-relativistic. It is possible, when
\begin{equation}
\tilde{\lambda}_{(T)} \left( 6q_1 + 6q_2 + q_3 \right) \left[1 + \nu_{(T)}\right] >1  + \tilde{\lambda}_{(S)}  \left[1 + \nu_{(S)}\right] \,.
\label{L76}
\end{equation}
We assume that the Stokes type force belongs to the class of friction forces, and thus it is characterized by positive parameter $\tilde{\lambda}_{(S)}$. Indeed, in the non-relativistic limit the spatial part of the force (\ref{13}) with $\nu_{(S)}=0$ takes the form
\begin{equation}
{\cal F}^{\alpha}_{(S)} \to  m_{(a)}\lambda_{(S)}[U^{\alpha} - \frac{1}{c} v^{\alpha} ]  \,,
\label{137}
\end{equation}
and if the particle velocity $v^{\alpha}$ $(\alpha = 1,2,3)$ exceeds the flow velocity $cU^{\alpha}$ the friction force takes negative value with positive parameter $\lambda_{(S)}$. The inequality (\ref{L76}) can be satisfied, if and only if its left-hand side is positive, i.e., the tidal force can be indicated as the accelerating one.

Clearly, when $\sigma_1=1$, the physical component of the particle momentum remains constant in the expanding Universe; this situation can be characterized, e.g., as a  "dynamic equilibrium". When $\sigma_1<1$, the particle momentum decreases.

We have to stress, that in this context there exists, formally speaking, the following model: since  the parameters in (\ref{L76}) depend on the sort index, for one part of particles $\sigma_1>1$, but for other particle sorts $\sigma_1<1$. This means that particles of the first subclass are accelerating and become ultra-relativistic; particles of the second subclass loss energy and become non-relativistic, asymptotically.

\subsubsection{Scale factor is of the power-law type }

Consider now the model with
\begin{equation}
\frac{a(t)}{a(t_0)} =  \left(\frac{t}{t_0} \right)^{\gamma} \,, \quad \frac{H(t)}{H(t_0)} = \left(\frac{t}{t_0} \right)^{-1} \,.
\label{pp1}
\end{equation}
Now we obtain that the behavior of the momentum is described also by the power law
\begin{equation}
{\cal P}(t) = {\cal P}(t_0)  \left(\frac{t}{t_0}\right)^{\gamma(\sigma_1-1)-\sigma_2}
\,.
\label{pp2}
\end{equation}
The momentum grows and the particle becomes ultra-relativistic asymptotically, when $\gamma(\sigma_1-1)>\sigma_2$, or in more details
\begin{equation}
\tilde{\lambda}_{(T)} (1+\nu_{(T)}) \left[3q_1(2\gamma-1) + 2q_2(3\gamma-2) + q_3(\gamma-1) \right] >
\gamma \left[\tilde{\lambda}_{(S)}(1+ \nu_{(S)}) + 1\right]
 \,.
\label{pp3}
\end{equation}
Similarly to the previous case, the energy of the particle decreases, when $\gamma(\sigma_1-1)<\sigma_2$, and it remains constant, if $\gamma(\sigma_1-1)=\sigma_2$.

\subsubsection{Particle behavior during the bounce regime}

In this paragraph we consider the model with the following scale factor and Hubble function
\begin{equation}
a(t) = a(t_*) e^{h t^2} \,, \quad H(t) = 2ht
 \,.
\label{qq1}
\end{equation}
The particle momentum evolves according to the law
\begin{equation}
{\cal P}(t) = {\cal P}(t_0)  e^{h (\sigma_1-1) (t^2-t^2_0)} \left(\frac{t}{t_0}\right)^{\sigma_2}
\,.
\label{qq2}
\end{equation}
Depending on the values of the parameters $\sigma_{1}-1$ and $\sigma_2$ five interesting situations appear.

1) When $\sigma_{1}>1$ and $\sigma_2>0$ the particle momentum (\ref{qq2}) grows monotonically; the particle is accelerating and becomes ultra-relativistic.

2) When $\sigma_{1}>1$ and $\sigma_2<0$ the particle momentum (\ref{qq2}) reaches the minimum at \\ $t=t_*= \sqrt{\frac{|\sigma_2|}{2h (\sigma_1-1)}}$, and then tends to infinity.

3) When $\sigma_{1}<1$ and $\sigma_2>0$ the particle momentum (\ref{qq2})  reaches the maximum at  \\  $t=t_*= \sqrt{\frac{\sigma_2}{2h (1-\sigma_1)}}$, and then decreases.

4) When $\sigma_{1}<1$ and $\sigma_2<0$ the particle momentum (\ref{qq2})  tends to zero monotonically.

5) When $\sigma_{1}=1$ and $\sigma_2=0$,  the particle momentum remains constant.

Fig. 1 illustrates these details of behavior.

\begin{figure}[H]
    \centering
    \begin{minipage}[b]{0.49\linewidth}
        \centering
        \includegraphics[width=1\linewidth, height=5cm]{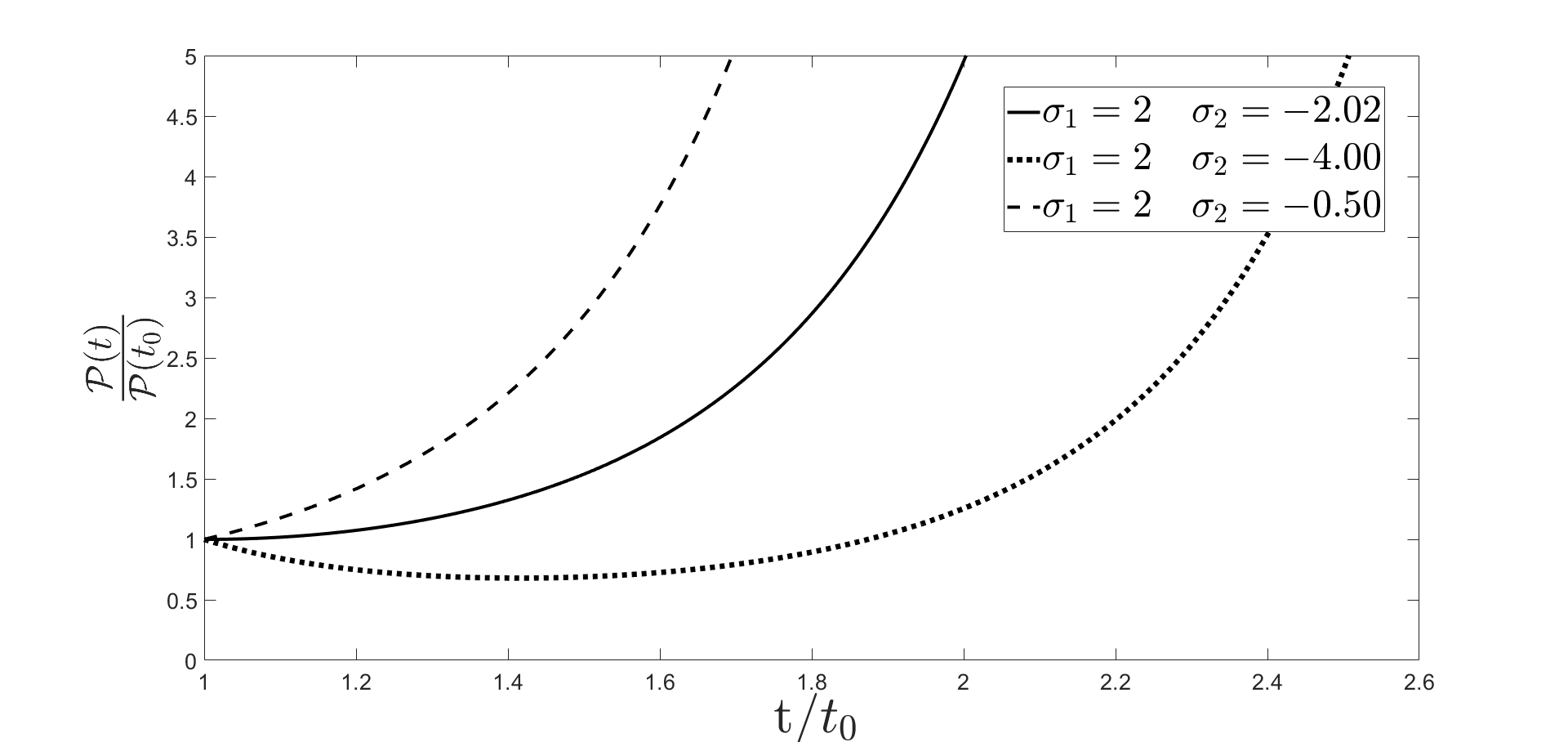}
        \caption*{a)}
    \end{minipage}
    \hfill
    \begin{minipage}[b]{0.49\linewidth}
        \centering
        \includegraphics[width=1\linewidth, height=5cm]{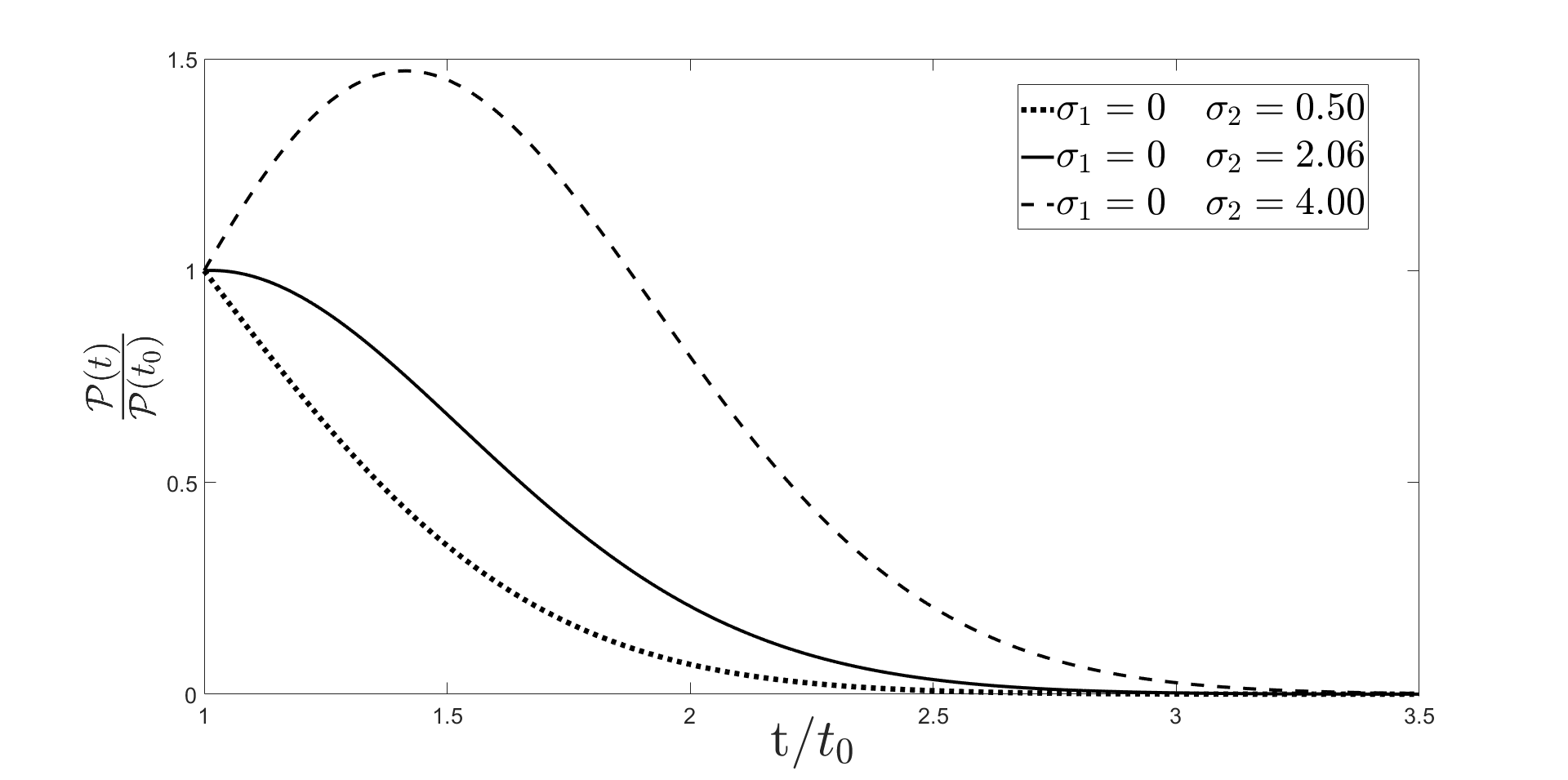}
        \caption*{b)}
    \end{minipage}
    \caption{
        Illustration of the behavior of the particle momentum for six sets of the dimensionless parameters $\sigma_{1}$ and $\sigma_{2}$.
        For illustration we put $h=1$. On the panel (a) we collect the increasing curves, which visualize the process of the ultra-relativistic particle creation;
        the curves presented on the panel (b) are the decreasing ones, they visualize the process of particle cooling.}
    \label{fig1}
\end{figure}

\subsubsection{Super-inflationary scenario}

Consider now the model with
\begin{equation}
\frac{a(t)}{a(t_0)} = e^{\nu \sinh(H_{*}(t-t_0)} \,, \quad \frac{H(t)}{H(t_0)} = \cosh{[H_{*}(t-t_0)]} \,.
\label{76}
\end{equation}
The law of evolution of the physical momentum component has the form
\begin{equation}
{\cal P}(t) = {\cal P}(t_0)\ e^{\nu (\sigma_1-1) \sinh(H_*(t-t_0))} \ (\cosh(H_*(t-t_0))^{\sigma_2} \,.
\label{77}
\end{equation}
Similarly to the previous case we highlight the same five details, but now  the extrema in the items 2) and 3) appear, if $|\sigma_2|>2\nu |\sigma_1-1|$, at the moment
\begin{equation}
t^* = t_0 + \frac{1}{H_*} {\rm Arsinh}\left\{-\frac{\sigma_2}{2 \nu (\sigma_1-1)} \pm \sqrt{\frac{\sigma^2_2}{4 \nu^2 (\sigma_1-1)^2}-1}\right\} \,.
\label{Sf77}
\end{equation}

\subsubsection{Quasi-periodic regime}

The last example relates to the case with
\begin{equation}
\frac{a(t)}{a(t_0)} = e^{\rho \sin{\Omega(t-t_0)}} \,, \quad \frac{H(t)}{H(t_0)} = \cos{\Omega(t-t_0)}  \,,
\label{94}
\end{equation}
\begin{equation}
{\cal P}(t) = {\cal P}(t_0) \ e^{\rho (\sigma_1-1) \sin{\Omega (t-t_0)} } \ (\cos{\Omega(t-t_0)})^{\sigma_2} \,.
\label{96}
\end{equation}
On the quasi-periodic platform the behavior of the particle momentum is also quasi-periodic without asymptotic growth; Fig. 2 illustrates this behavior.

\begin{figure}[H]
\centerline{\includegraphics[width=0.5 \linewidth,height=5cm]{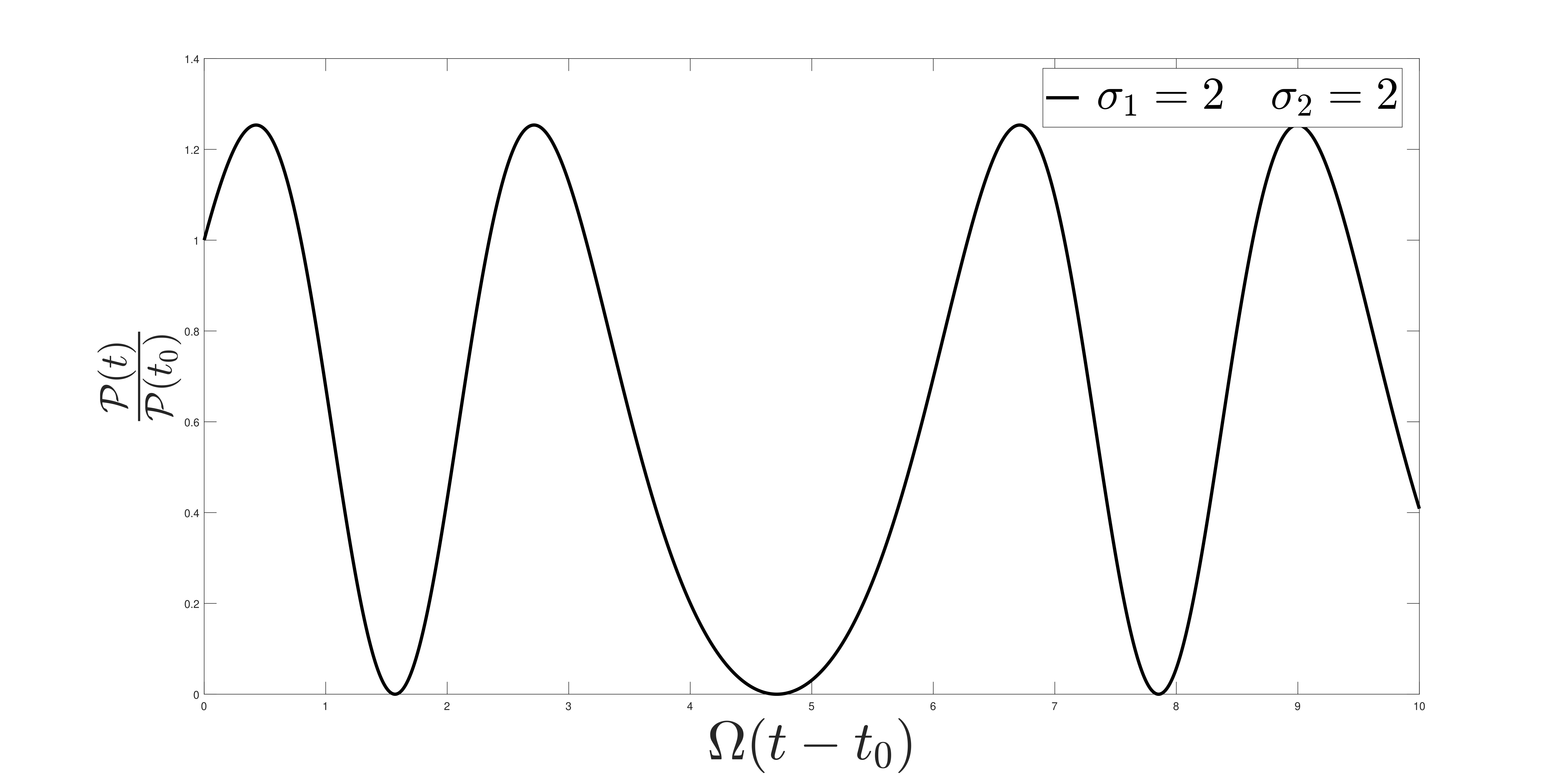}}
\caption{Illustration of the behavior of the particle momentum for the quasi-periodic space-time platform. For illustration we put $\rho {=} 1$.}
\label{fig2}
\end{figure}

\subsection{Axion field is not constant}

Looking at the formula (\ref{L87}) we can conclude that the typical behavior of the physical component of the particle momentum can be described by the formula:
\begin{equation}
{\cal P}(t) = {\cal P}(t_0) \left(\frac{a(t_0)}{a(t)}\right) \exp\left\{ \lambda_{(G)}  \left[\cos{\phi}(t) - \cos{\phi}(t_0)\right]\left[1+ \frac12 \nu_{(G)}\left(\cos{\phi}(t) + \cos{\phi}(t_0)\right) \right] \right\}  \,.
\label{A11}
\end{equation}
 Clearly, the exponential part of this formula is periodic function; the multiplier  $\frac{a(t_0)}{a(t)}$ describes the tendency of decreasing of the particle momentum.
In order to complete the study of this model, we consider several examples of the behavior of the axion field, which enters the arguments of the trigonometric functions.

\subsubsection{Standard inflation}

Studying the model of standard inflation, we assume that, first, the scale factor is presented by the formula $a(t)=a(t_0) e^{H_0 (t-t_0)}$ with $H_0 = {\rm const}$, second, $\phi(t_0)=0$, third, $\phi(t)$ is small, fourth, $1+{\cal A}(H)>0$. Then we obtain the evolutionary equation for the axion field in the following form
\begin{equation}
\ddot{\phi} + 3H_0\dot{\phi} +\frac{m^2_A}{1+ {\cal A}(H_0)}\phi = 0 \,.
\label{115}
\end{equation}
Depending on the value of the axionic mass $m_A$, three variants of solutions can be obtained.

1. If $m_{A}< \frac{3H_0}{2} \sqrt{1+{\cal A}(H_0)}$, the solution is
\begin{equation}
\phi(t) = \frac{\dot{\phi}(t_0)}{\Gamma} e^{-\frac{3H_0}{2}(t-t_0)} \sinh{\Gamma (t-t_0)}  \,, \quad \Gamma = \sqrt{\frac94 H^2_0 - \frac{m^2_{A}}{1+{\cal A}(H_0)}} \,.
\label{SPS1}
\end{equation}
In order to visualize the behavior of the particle momentum (\ref{A11})  for the axion field described by (\ref{SPS1}), we introduce two auxiliary dimensionless parameters $\Lambda {=} \frac{2m_{A}}{3H_0 \sqrt{1+{\cal A}(H_0)}}$ and $\Theta {=} \frac{\dot{\phi}(t_0)}{H_0}$ (see the legends on Fig. 3 and 4).

\begin{figure}[H]
\hfill
\begin{minipage}[b]{0.49 \linewidth}
\center{\includegraphics[width=1\linewidth, height=5cm ]{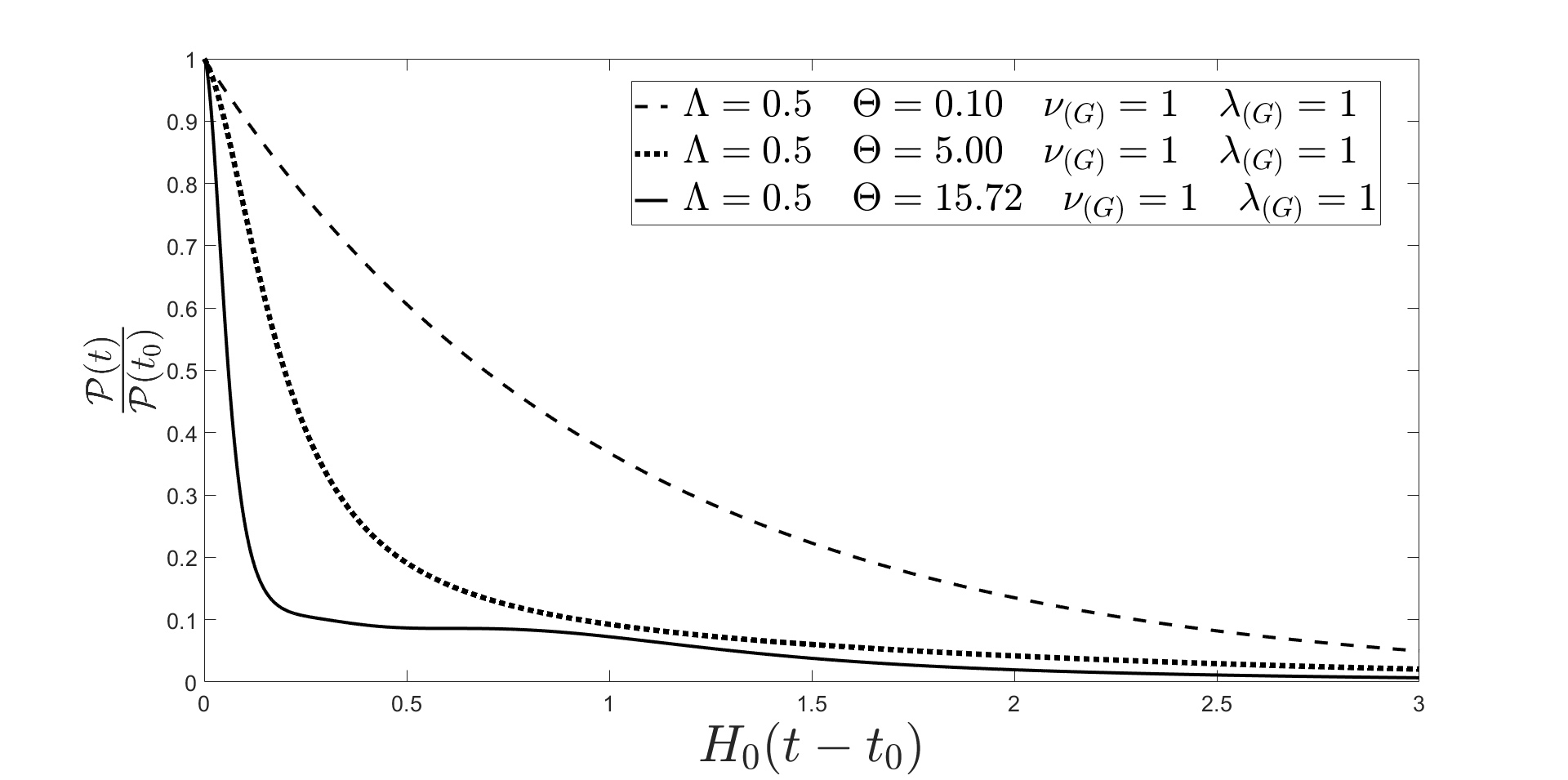}} a) \\
\end{minipage}
\hfill
\begin{minipage}[b]{0.49\linewidth}
\center{\includegraphics[width=1 \linewidth , height=5cm]{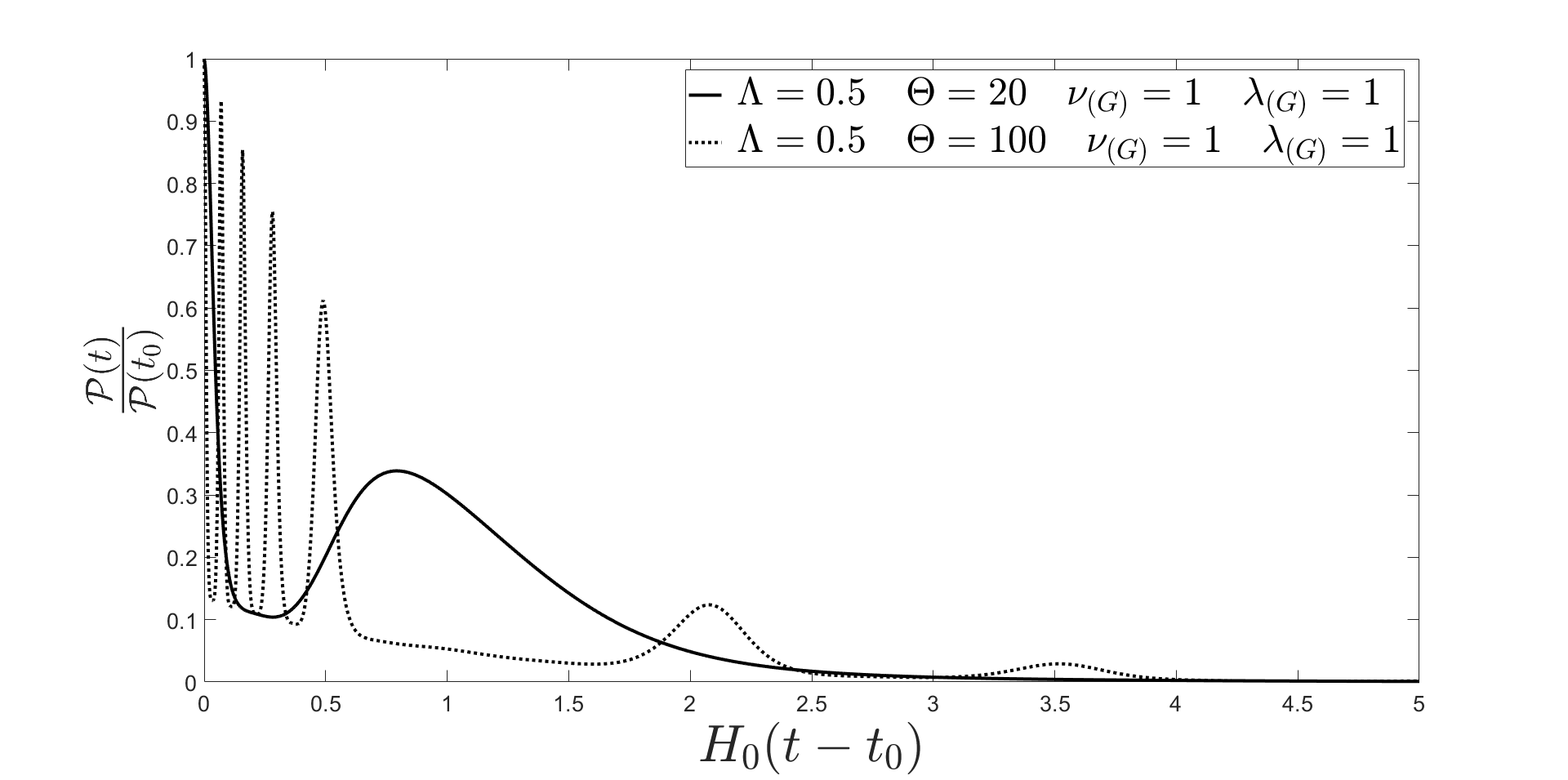}} b) \\
\end{minipage}
\hfill\text{}
\caption{Illustration of the formula (\ref{A11}) for the case (\ref{SPS1}), when only one parameter $\Theta$ is varying. On the panel a) the third curve with $\Theta  {=} \Theta_* {\approx} 15.72$ plays the role of some separatrix: when $\Theta {<} \Theta_*$, extrema do not appear. On the panel b) we present the illustrations with $\Theta{>}\Theta_*$; clearly, these curves have minima and maxima.}
\label{fig3}
\end{figure}

\begin{figure}[H]
\hfill
\begin{minipage}[b]{0.49 \linewidth}
\center{\includegraphics[width=1\linewidth, height=5cm ]{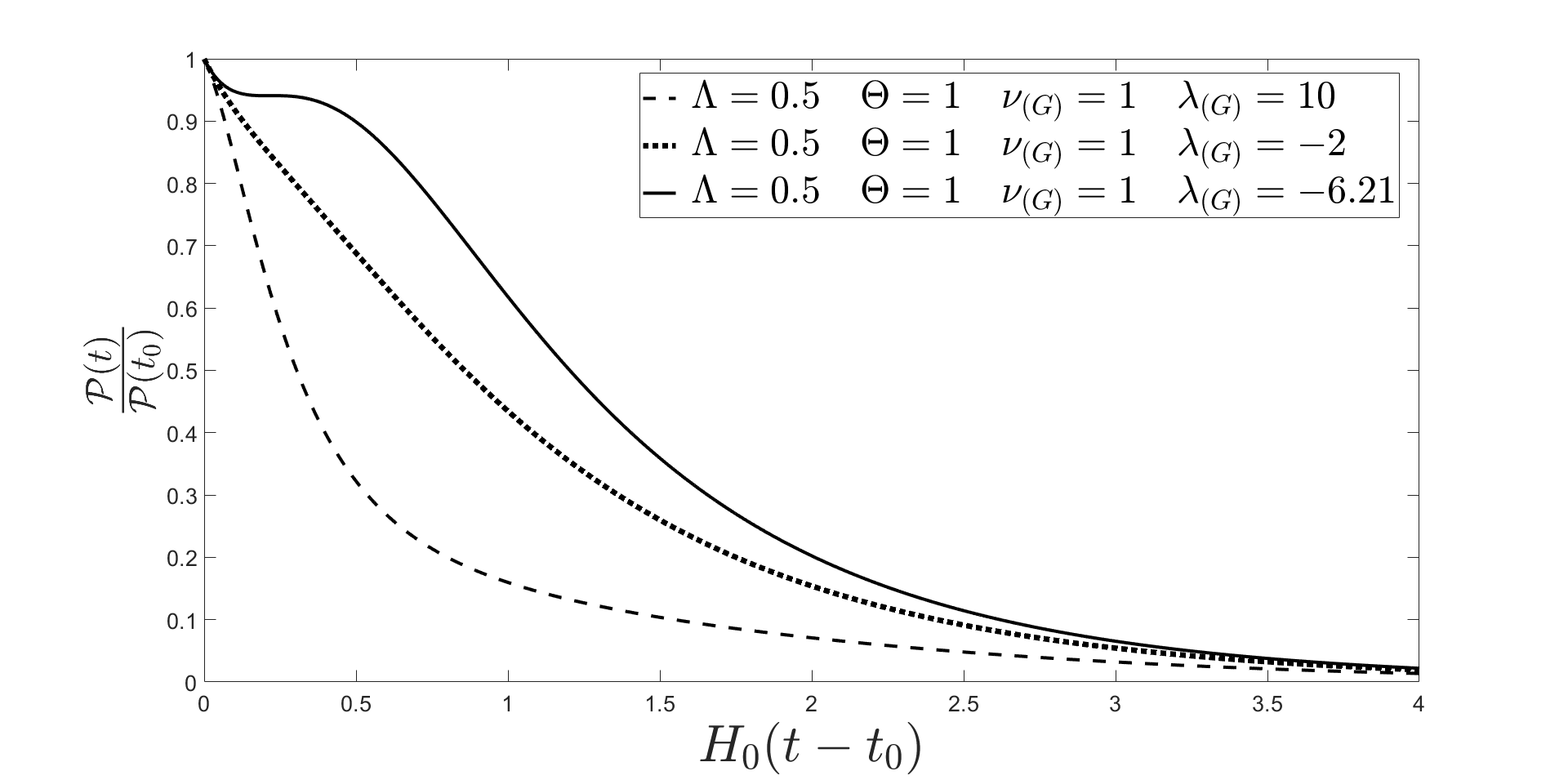}} a) \\
\end{minipage}
\hfill
\begin{minipage}[b]{0.49\linewidth}
\center{\includegraphics[width=1 \linewidth , height=5cm]{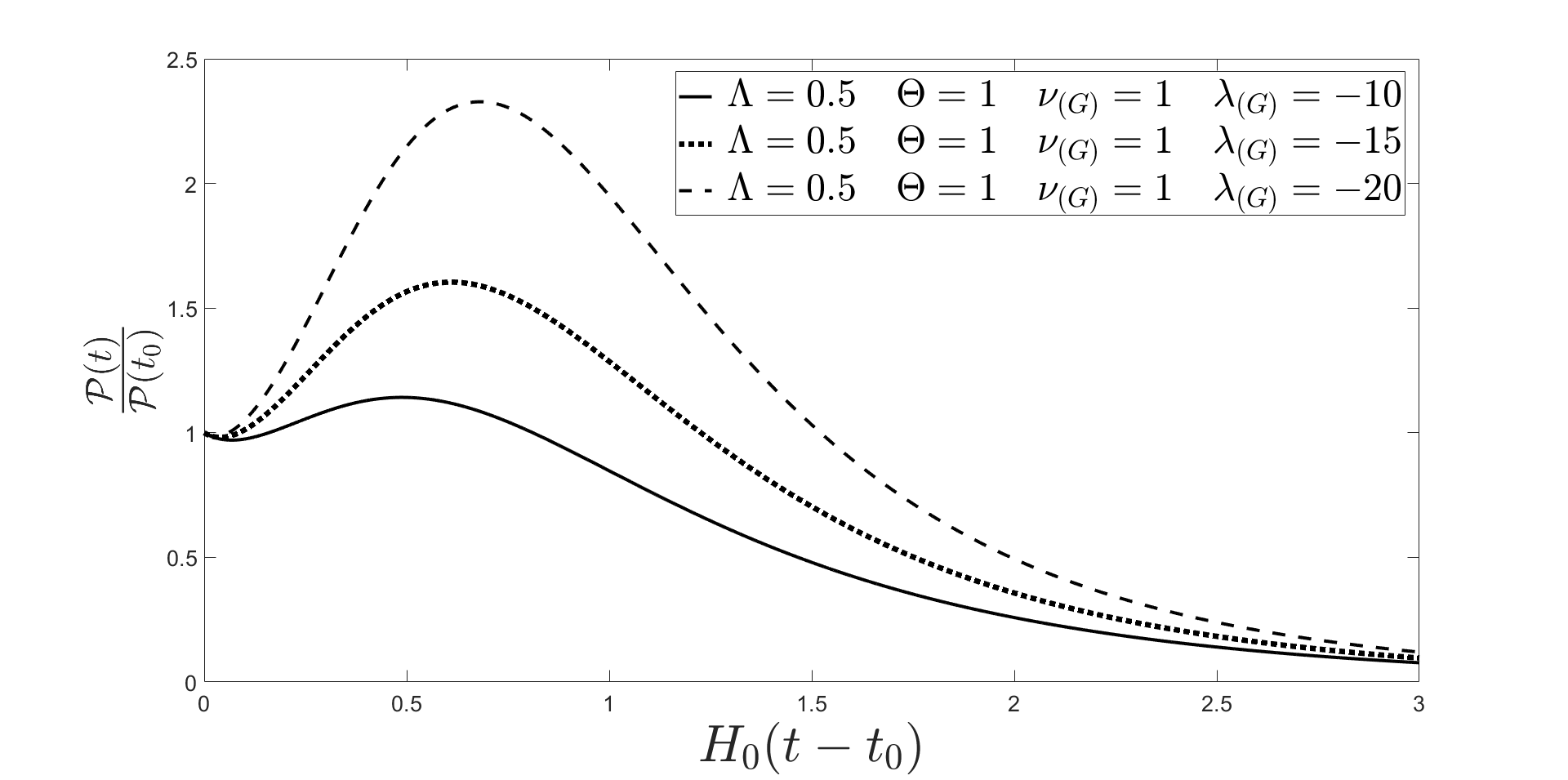}} b) \\
\end{minipage}
\hfill\text{}
\caption{Illustration of the formula (\ref{A11}) for the case (\ref{SPS1}), when only one parameter $\lambda_{(G)}$ is varying and the value of $\Theta$ is fixed by $\Theta{=}1{<}\Theta_*$. On the panel a) the upper curve with $\lambda_{(G)}{=} \lambda_{(G)}^{*} {\approx} -6.21$ plays the role of separatrix: when $\lambda_{(G)} {>} \lambda_{(G)}^{*}$, extrema do not appear. On the panel b) we present the illustrations with $\lambda_{(G)} {<} \lambda_{(G)}^{*}$; these curves have maxima.}
\label{fig4}
\end{figure}

2. If $m_{A}> \frac{3H_0}{2} \sqrt{1+{\cal A}(H_0)}$, the solution $\phi(t)$ is quasi-periodic
\begin{equation}
\phi(t) = \frac{\dot{\phi}(t_0)}{\Omega_0} e^{-\frac{3H_0}{2}(t-t_0)} \sin{\Omega_0 (t-t_0)}  \,, \quad \Omega_0 = \sqrt{-\frac94 H^2_0 + \frac{m^2_{A}}{1+{\cal A}(H_0)}} \,.
\label{SPS2}
\end{equation}
The behavior of the particle momentum can be now indicated as quasi-double-periodic; the illustration of this case is presented on the panel a) of the Fig. 5.

3. If $m_{A}= \frac{3H_0}{2} \sqrt{1+{\cal A}(H_0)}$, the solution for the axion field takes the form
\begin{equation}
\phi(t) = \dot{\phi}(t_0) (t-t_0) e^{-\frac{3H_0}{2}(t-t_0)}   \,.
\label{SPS3}
\end{equation}
The behavior of the particle momentum is illustrated on the panel b) of the Fig. 5.

\begin{figure}[H]
\hfill
\begin{minipage}[b]{0.49 \linewidth}
\center{\includegraphics[width=1\linewidth, height=5cm ]{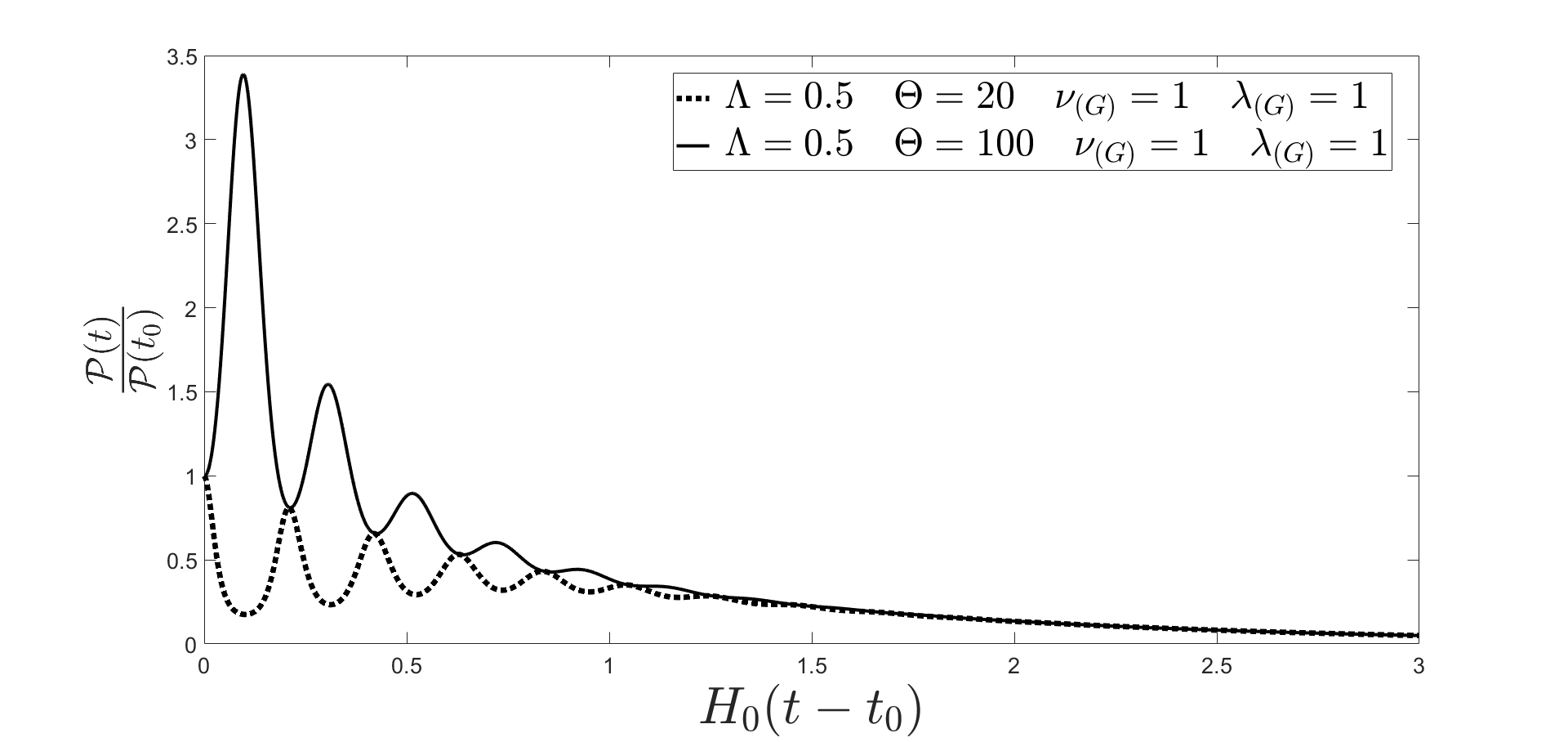}} a) \\
\end{minipage}
\hfill
\begin{minipage}[b]{0.49\linewidth}
\center{\includegraphics[width=1 \linewidth , height=5cm]{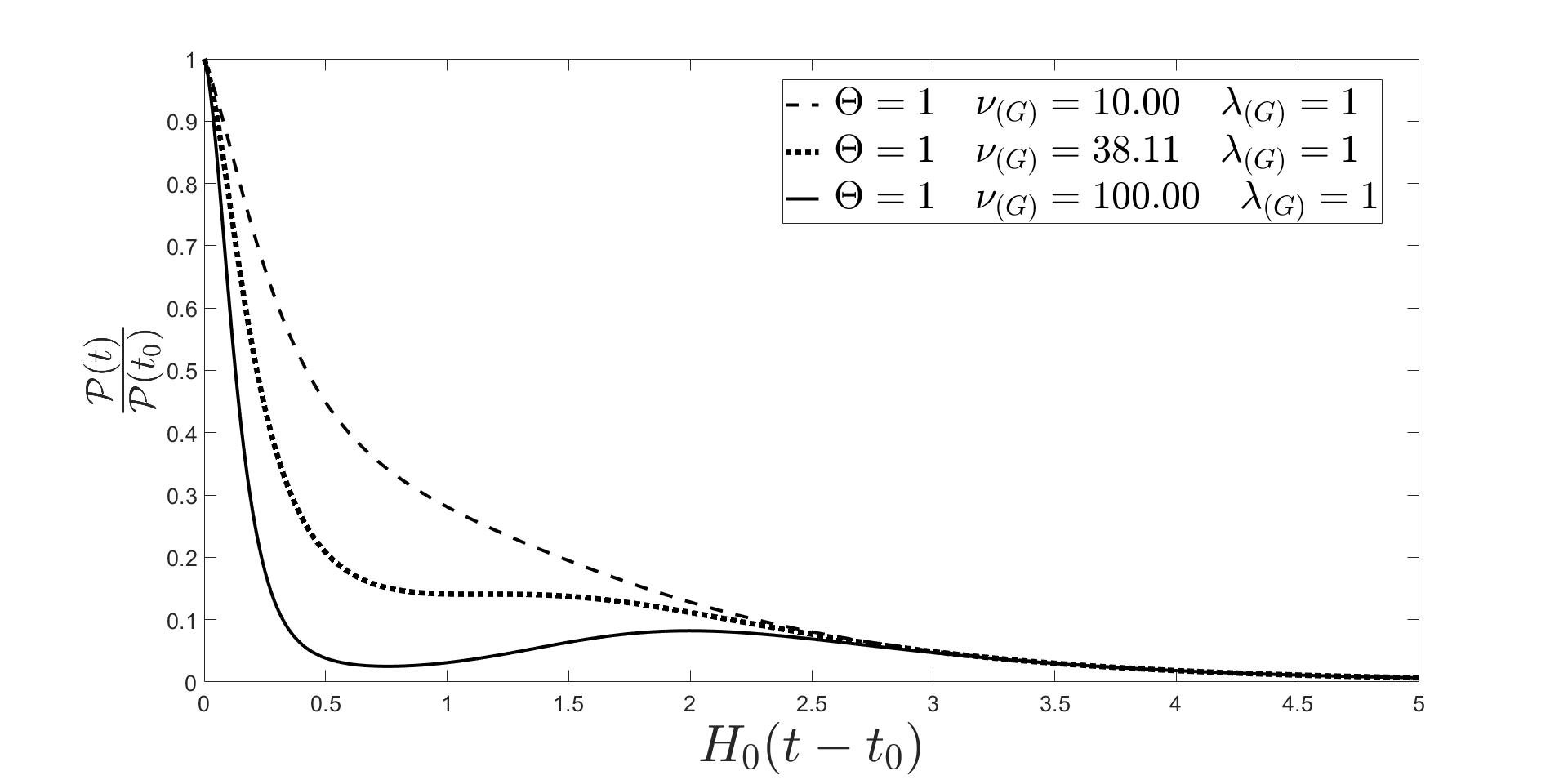}} b) \\
\end{minipage}
\hfill\text{}
\caption{Illustration of the behavior of the particle momentum: the panel a) corresponds to the formula (\ref{SPS2}) and panel (b) to the formula (\ref{SPS3}).}
\label{fig5}
\end{figure}

\subsubsection{Power-law regime of evolution}

If $a(t) = a(t_0) \left(\frac{t}{t_0}\right)^{\gamma}$ and $H=\frac{\gamma}{t}$, we assume that ${\cal A}=0$, and illustrate the model by the solution to the equation
\begin{equation}
\ddot{\phi} + \frac{3 \gamma}{t} \dot{\phi} + m^2_{A} \phi = 0  \,.
\label{SPS4}
\end{equation}
Now the axion field can be presented in terms of Bessel functions
\begin{equation}
\phi(t) = \left(\frac{t}{t_0} \right)^{\frac{1-3\gamma}{2}} \left[C_1 J_{\nu_*}(m_{A}t) + C_2 Y_{\nu_*}(m_{A}t)\right] \,, \quad \nu^2_*=\frac{9\gamma^2-1}{4}  \,.
\label{SPS5}
\end{equation}
Here $J_{\nu_*}(m_{A}t)$ and $Y_{\nu_*}(m_{A}t)$ are the Bessel functions of the first and second kinds, respectively; the constants of integration $C_1$,  $C_2$ can be found from the initial conditions, they depend on the choice of the sign of the index $\nu_*$. An illustration of the behavior of the particle momentum for this case is presented on the panel a) of the Fig. 6.

\subsubsection{Evolution of massless axions}

For the last illustration we consider the model with
\begin{equation}
m_{A} =0 \,, \quad \frac{1 + {\cal A}(H(t_0))}{1 + {\cal A}(H(t))} = 3 \frac{H(t)}{H(t_0)}  \,.
\label{SPS6}
\end{equation}
Then the solution to the equation (\ref{M55}) is
\begin{equation}
\phi(t) = \phi(t_0) + \frac{\dot{\phi}(t_0)}{H(t_0)} \left[1- \left(\frac{a(t_0)}{a(t)} \right)^3 \right] \,.
\label{SPS61}
\end{equation}
Respectively, during the super-inflation stage the axion field behaves as
\begin{equation}
\phi(t) = \phi(t_0) + \frac{\dot{\phi}(t_0)}{H(t_0)} \left[1- e^{-3 \nu \sinh{H_{*}(t-t_0)}} \right] \,,
\label{SPS7}
\end{equation}
and the quasi-periodic regime is characterized by the behavior
\begin{equation}
\phi(t) = \phi(t_0) + \frac{\dot{\phi}(t_0)}{H(t_0)} \left[1- e^{-3 \rho \sin{\Omega(t-t_0)}} \right] \,.
\label{SPS71}
\end{equation}
An illustration of the behavior of the particle momentum for the case (\ref{SPS71}) is presented on the panel b) of the Fig. 6.

\begin{figure}[H]
\hfill
\begin{minipage}[b]{0.49 \linewidth}
\center{\includegraphics[width=1\linewidth, height=5cm ]{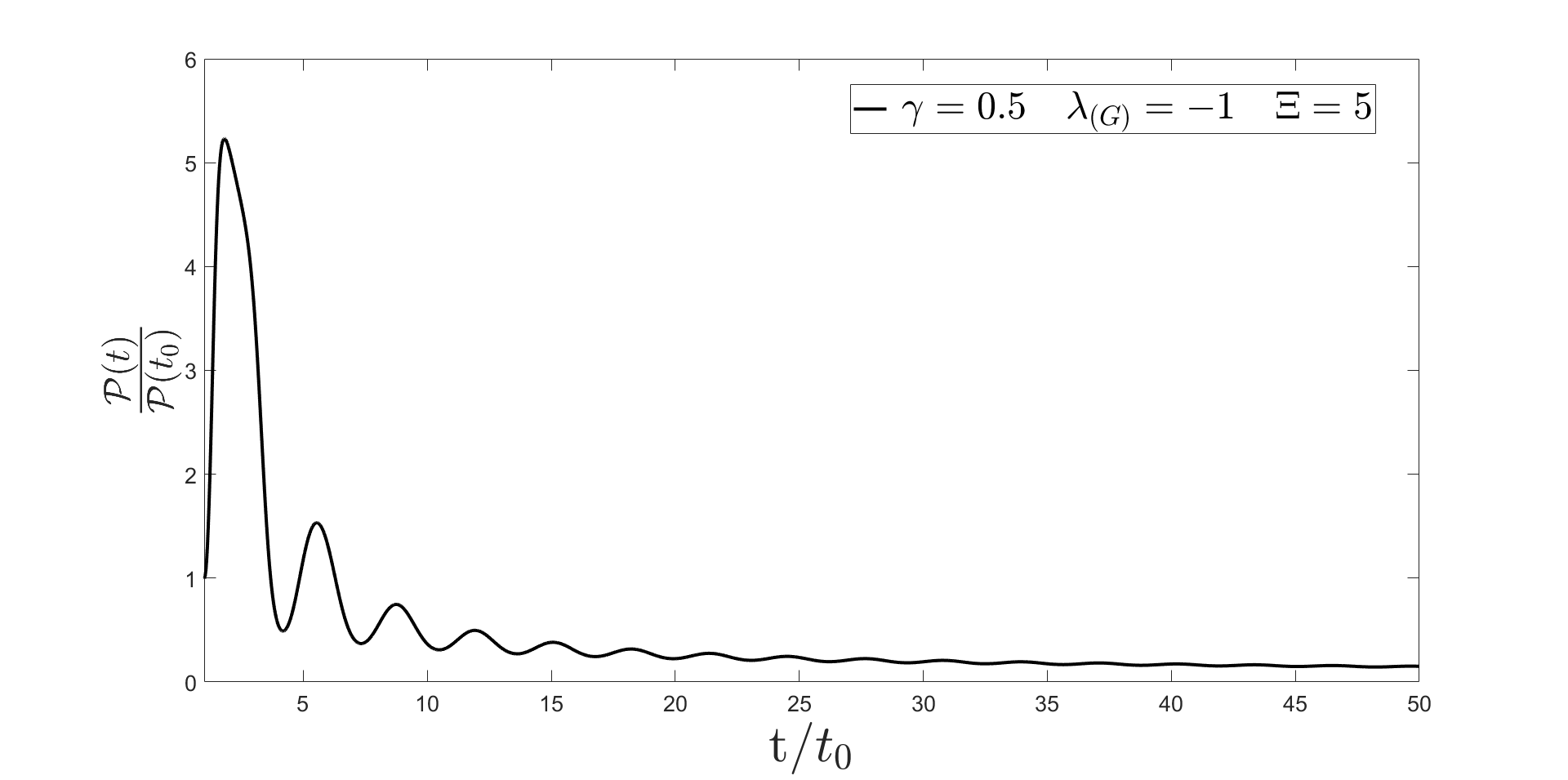}} a) \\
\end{minipage}
\hfill
\begin{minipage}[b]{0.49\linewidth}
\center{\includegraphics[width=1 \linewidth , height=5cm]{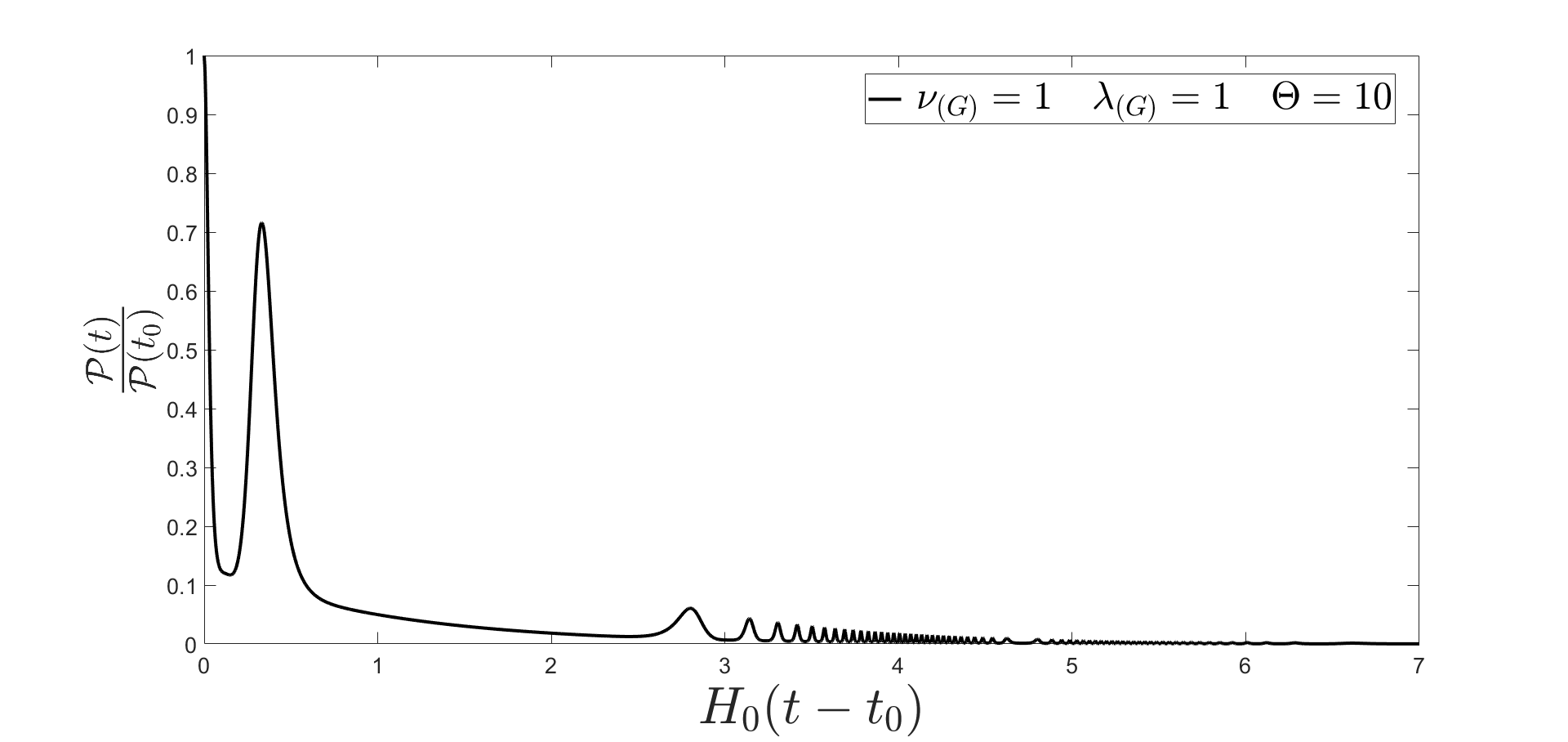}} b) \\
\end{minipage}
\hfill\text{}
\caption{The panel a) illustrates the behavior of the particle momentum for the axion field described by the formula (\ref{SPS5}) with $\phi(t_0){=}0$ and $m_A t_0 {=} 1$. The panel (b) visualizes the case related to the formula (\ref{SPS71}) with $\phi(t_0){=}0$ and $\nu {=} 1$. Here $\Xi {=} \frac{\dot{\phi}(t_0)}{m_{A}}$ and $\Theta {=} \frac{\dot{\phi}(t_0)}{H_0}.$}
\label{fig:RC2}
\end{figure}

\section{Conclusions}

We have reconstructed the complete functions $f_{(a)}(x,p)$, which describe the distribution of the relativistic particles of the sort $a$ in the Phase Space.
These distribution functions have the structure $f_{(a)}(K_1,K_2,K_3,K_4,K_5,K_6)\delta(K_0-m^2_{(a)}c^2) $, where $K_0$, $K_1$,...$K_6$ are the so-called integrals of motion, the exact solutions to the characteristic equations. The analysis of behavior of the physical component of the particle momentum ${\cal P}(t)$ has shown that the curvature induced force, linear in the Riemann tensor, can divide the particle ensemble into two sub-groups, if the guiding parameters $\lambda_{(\rm T)}$, $\lambda_{(\rm S)}$, etc., depend on the sort of particles. Particles of the first sub-group loss the energy and momentum monotonically, and become non-relativistic asymptotically. These particles can be characterized by the temperature $T_{(\rm Cold)}$, for which $m_{(a)}c^2 >> k_{B} T_{(\rm Cold)}$. Particles of the second sub-group acquire energy from the curvature force, they can be attributed to the class of hot particles. Depending on the values of the guiding parameters of the model, these particles can become ultra-relativistic asymptotically (see, e.g., the panel a) of Fig.1), or can reach some maximal energy (see, e.g., the panel b) of Fig. 3, panel b) of Fig. 4, panel a) of Fig. 5). These particles can be characterized by the temperature $T_{(\rm Hot)}$. One can assume that $T_{(\rm Hot)}>>T_{(\rm Cold)}$. In other words, the curvature force can turn the multi-component plasma into the so-called {\it two-temperature plasma}. Since the plasma particles are assumed to be electrically charged, the {\it plasma instabilities} can occur in the axion-aether-active plasma, if some internal or external perturbations appear in plasma.
For instance, if the gravitational waves travel through the two-temperature plasma, they produce longitudinal electric currents and allow us to speak about symptoms of plasma instability \cite{BaIg,Ba82}. The problem of plasma instabilities in the axion-aether-active plasma will be the topic of the third part of our work.

\end{document}